\documentclass[a4paper,fleqn]{cas-dc}

\usepackage[numbers,sort&compress]{natbib}

\def\tsc#1{\csdef{#1}{\textsc{\lowercase{#1}}\xspace}}
\tsc{WGM}
\tsc{QE}
\tsc{EP}
\tsc{PMS}
\tsc{BEC}
\tsc{DE}

\begin{document}
\let\WriteBookmarks\relax
\def\floatpagepagefraction{1}
\def\textpagefraction{.001}

\shorttitle{Brillouin scattering induces optomechanical entanglement}    

\shortauthors{}  

\title [mode = title]{Optomechanical entanglement induced by backward stimulated Brillouin scattering}  

\tnotemark[1] 

\tnotetext[1]{} 

%

\author[1]{P. Djorwé}

\cormark[1]


\ead{djorwepp@gmail.com}

\ead[url]{}

\credit{Conceptualization, investigation, analysis and writing}

\affiliation[1]{organization={Department of Physics, Faculty of Science, University of Ngaoundere},
            city={Ngaoundere},
         citysep={}, 
            postcode={454}, 
            country={Cameroon}}

\author[2]{A.H. Abdel-Aty}
%
%
 \ead{amabdelaty@ub.edu.sa}
%
%
\credit{Discussions, Methodology and supervision}
 \affiliation[1]{organization={Department of Physics, College of Sciences, University of Bisha},
          city={Bisha},
            citysep={}, 
            postcode={61922}, 
          country={Saudi Arabia}}

\author[3]{K.S. Nisar}
%
%
 \ead{n.sooppy@psau.edu.sa}
%
%
\credit{Discussions, supervision}
 \affiliation[3]{organization={Department of Mathematics, College of Science and Humanities in Alkharj, Prince Sattam Bin Abdulaziz University},
          city={Alkharj},
            citysep={}, 
            postcode={11942}, 
          country={Saudi Arabia}}

\author[3]{S. G. Nana Engo}
%
%
 \ead{serge.nana-engo@facsciences-uy1.cm}
%
%
\credit{Methodology, Discussions, supervision}
 \affiliation[3]{organization={Department of Physics, Faculty of Science, University of Yaounde I},
          city={Yaounde },
            citysep={}, 
            postcode={812}, 
          country={Cameroon}}


\begin{abstract}
We propose a scheme to generate robust optomechanical entanglement. This scheme is based on a Backward Stimulated Brillouin Scattering (BSBS) process, which is hosted within an optomechanical structure. Our benchmark system consists of an acoustic (mechanical) mode coupled to two optical modes through the BSBS (radiation pressure) process. For a moderate values of the effective mechanical coupling, the BSBS induces a relatively weak entanglement. This entanglement is greatly enhanced, for at least up to one order of magnitude, when the mechanical coupling strength is strong enough. The generated entanglement is robust enough against thermal fluctuation. Our work provides a new scheme for entanglement generation based on BSBS effect, and can be extended to microwaves and hybrid optomechanical structures. Such a generated entangled states can be used for quantum information processing, quantum sensing, and quantum computing.   
\end{abstract}

\begin{highlights}
\item The generated entanglement depends on acoustic parameters from specific threshold;
\item The generated entanglement is greatly enhanced for at least up to one order of magnitude;
\item The generated entanglement survives for thermal phonon occupancy up to $n_{th}=200$.
\end{highlights}


\maketitle


\section{Introduction}\label{intro}

Optomechanical interaction, coupling between electromagnetic fields and mechanical objects, has recently fostered a number of interesting phenomena ranging from classical \cite{Walter_2016,Pokharel_2022,Foulla_2017,Djor.2022} to quantum \cite{Aspelmeyer_2010,Lemonde.2016,Djor.2012} regimes. In the classical regime, intriguing behavior such as collective phenomena \cite{Li_2022,Colombano_2019,Djorwe.2020}, nonlinear dynamic \cite{Roque.2020,Xu_2024} and chaos \cite{Navarro.2017,Zhu_2023,Stella_2023} have been uncovered. These phenomena have been used to improve applications such as communication schemes based-synchronization \cite{Rodrigues_2021,Berra_2023}, random number generation \cite{Madiot_2022}, and encryption schemes based-chaos \cite{Wu_2019}. In the quantum regime, a plethora of quantum phenomena have been induced  to enhance a number of quantum technologies \cite{Barzanjeh_2021}. Among these quantum phenomena we can list squeezed states \cite{Qin_2022,Banerjee_2023}, entangled states \cite{Riedinger_2018,Kotler_2021}, quantum correlations \cite{Riedin.2016,Purdy_2017}, and quantum synchronization \cite{Mari_2013,Garg_2023}.      

Quantum entanglement constitutes a key ingredient for quantum technologies including quantum computing \cite{Zidan_2021}, quantum information processing \cite{Meher_2022},  quantum sensing \cite{Xia_2023,Brady_2023}, and quantum teleportation \cite{Fiaschi_2021} among others. However, quantum entanglement resources are limited by decoherence and quantum fluctuations. Owing to their pivotal role in numerous quantum applications, different schemes have been proposed to generate robust entangled states that resist against decoherence and quantum fluctuations. For this purpose, nonlinear phenomena such as cross-Kerr effect, parametric  amplification, and Duffing nonlinearity have been used to induce quantum entanglement in optomechanical systems \cite{Zhang_2017,Cha_2017}. Recently, Exceptional Points (EPs), which are Non-Hermitian degeneracies, have been equally proposed to engineer robust quantum entanglement in optomechanics \cite{Chakrabo.2019}. Moreover, sophisticated schemes based on dark mode and symmetry breaking have been also put forward to synthesize large amount and robust entanglement in optomechanics \cite{Lai.2022,Lai_2022}. Beside these techniques, there are a number of interesting ongoing optomechanical activities involving Brillouin scattering processes \cite{Bahl_2013,He_2020,Doeleman_2023,Wang_2024}. In our knowledge, there is a lack of schemes for entanglement generation based on this Brillouin scattering process in the literature.      

In this letter, we propose a scheme to generate robust quantum entanglement in optomechanical system, whose scheme is based on Backward Stimulated Brillouin Scattering (BSBS) process. Our benchmark system consists of an acoustic (mechanical) mode coupled to two optical modes through the BSBS process (radiation pressure). We have found that i) a moderate values of the effective mechanical coupling induces relatively weak entanglement that depends on acoustic parameters which are appropriately chosen from specific threshold, and ii) a strong enough mechanical coupling strength greatly enhances entanglement for at least up to one order of magnitude. Moreover, we have shown that the generated entanglement is robust against thermal fluctuation, and this entanglement survives for thermal phonon occupancy up to $n_{th}=200$. Our work paves a way toward a new scheme  that can be used to generate stable and robust entanglement based on BSBS effect. This scheme can be extended to microwave systems and other hybrid optomechanical structures. Such a generated entangled states can be useful for quantum information processing, quantum sensing and metrology, and quantum computing. 

The rest of our work is organized as follows. \autoref{sec:model} provides the dynamical equations and derives the analytical expressions involved in our proposal. The steady-state quantum entanglement together with the important results are presented throughout \autoref{sec:entag}. Our work in concluded in \autoref{sec:concl}.

\section{Model and dynamical equations} \label{sec:model}
Our benchmark system consists of an acoustic mode $b_a$ that couples to two optical modes through the BSBS process, and a mechanical oscillator $b_m$ that couples to the same optical modes through the standard radiation pressure coupling. The Hamiltonian of such a system is given by ($\hbar=1$):
\begin{equation}\label{eq:eq1}
 H =H_{\rm{0}}+H_{\rm{OM}}+H_{\rm{BSBS}}+H_{\rm{drive}}, 
 \end{equation}
where
\begin{eqnarray}  \label{eq:eq2}
H_{\rm{0}}&:=&\sum_{j=1,2} \omega_{c_j} a_j^\dagger  a_j + \omega_a b_a^\dagger b_a + \omega_m b_m^\dagger b_m, \\
H_{\rm{OM}}&:=&\sum_{j=1,2}g_{m_j} a_j^\dagger a_j   (b_m +b_m^\dagger), \\
H_{\rm{BSBS}}&:=&- g_a( a_1^\dagger a_2 b_a+ a_1 a_2^\dagger b_a^\dagger),\\
H_{\rm{drive}}&:=&\sum_{j=1,2}iE_j(a_j^{\dagger }e^{-i\omega_{p_j} t} - a_je^{i\omega_{p_j} t}).
\end{eqnarray} 
In the above Hamiltonian, the operators $a_j$ capture the optical fields involved, and $H_{\rm{0}}$ stand for the free Hamiltonian of the whole system. The term $H_{\rm{OM}}$ represents the optomechanical interaction between the optical field ($a_j$) and the mechanical mode ($b_m$) through the single-photon couplings $g_{m_j}$. The term $H_{\rm{BSBS}}$ stands for the triply resonant phonon-photon interactions triggered via the $\rm{BSBS}$  process. $H_{\rm{drive}}$ is the driving field's Hamiltonian, where $E_j$ and $\omega_{p_j}$ are the amplitude and frequency of the $j^{th}$ field. For now on, we assume $g_{m_j}\equiv g_m$, and we set the optical cavity frequency as $\omega_{c_j}$ while the mechanical (acoustic) frequency is $\omega_m$ ($\omega_a$).

In the frame rotating at $H_r=\omega_{p_1}a_1^\dagger a_1 +\omega_{p_2} a_2^\dagger a_2 +({\omega_{p_1}-\omega_{p_2}})b_a^\dagger b_a $, the Hamiltonian in \autoref{eq:eq1} becomes
\begin{eqnarray}\label{eq:eq3}
H'&=- \Delta_1 a_1^\dagger  a_1 + \Delta_a b_a^\dagger b_a+\omega_m b_m^\dagger b_m- g_{m} a_1^\dagger a_1 (b_m +b_m^\dagger)\nonumber \\&+iE_1(a_1^{\dagger } - a_1) - G_a( a_1^\dagger b_a+ a_1 b_a^\dagger),
\end{eqnarray}
where we have defined $\Delta_1=\omega_{p_1}-\omega_{c_1}$, and $\Delta_a=\omega_a+\omega_{p_2}-\omega_{p_1}$. Here ${G_a}=g_a \alpha_2$, where $\alpha_2$ is the steady-state of the control optical mode $a_2$, which has been treated classically 

\begin{figure}[tbh]
  \begin{center}
  \resizebox{0.4\textwidth}{!}{
  \includegraphics{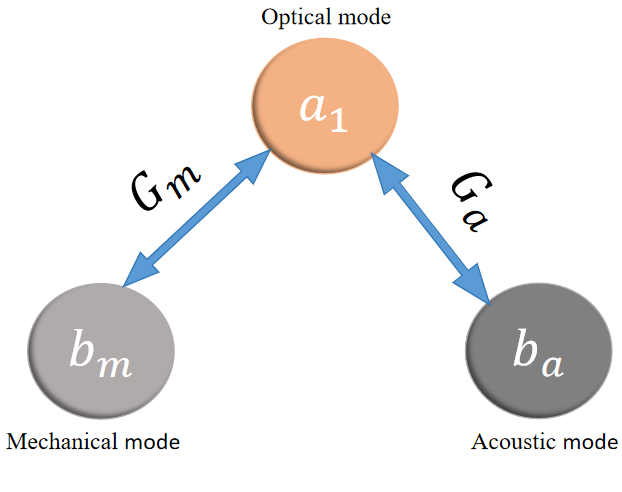}}
  \end{center}
\caption{Sketch of our linearized three mode optomechanical system. The mode $a_1$ is coupled to the acoustic (mechanical) mode $b_a$ ($b_m$) through the Brillouin (optomechanical) coupling $G_a$ ($G_a$), which are induced from  electrostrictive (radiation pressure) force. The phonon-phonon hopping rate $J_m$ is modulated by the phase $\theta$.}
\label{fig:Fig1}
\end{figure}

By using the well-known linearization process in optomechanics, we  can linearize \autoref{eq:eq3}, and our resulting three modes optomechanical system can be sketched as depicted in \autoref{fig:Fig1}. This linearized Hamiltonian leads to the following dynamical set of equations for the fluctuation operators,  
\begin{equation}\label{eq:fluc}
\begin{cases}
\delta\dot{a}_1 &= \left(i\tilde{\Delta} - \frac{\kappa}{2}\right) \delta a_1 + iG_c({\delta}b_{m}^{\dagger} +{\delta}b_m) +i{G_b} \delta b_a \\&+\sqrt{\kappa}a_1^{in}\\ 
\delta\dot{b}_a &= - ({\frac{\gamma_a}{2}} + i{\Delta_a})\delta{b}_a  - i{J_{m}}e^{i\theta}\delta{b}_m +i{G_a} \delta a_1 \\&+\sqrt{\gamma_a}b_{a}^{in} \\
\delta\dot{b}_m &= -({\frac{\gamma_m}{2}}+ i {\omega_m})\delta{b}_m   -i{J_{m}}e^{-i\theta}\delta{b}_a + i(G_m^*\delta a_1 \\&+ G_m\delta a_1^\dagger ) + \sqrt{\gamma_m}b_{m}^{in}, 
\end{cases}
\end{equation}
where we have accounted the optical ($\kappa$), mechanical ($\gamma_m$) and acoustic ($\gamma_a$) dissipation rates, respectively. Moreover, we have defined the effective detuning $\tilde{\Delta}=\Delta-2g_m\rm Re(\beta_m)$, where $\beta_m$ is the steady-state of the mechanical mode. In \autoref{eq:fluc}, $a_1^{in}$, $b_a^{in}$ and $b_m^{in}$ are zero-mean noise operators characterized by the following auto-correlation functions,
\begin{eqnarray}\label{eq:noise}
&\langle a_1^{in}(t)a_1^{in\dagger}(t') \rangle =\delta(t-t'), \hspace*{0.5cm} \langle a_1^{in\dagger}(t)a_1^{in}(t') \rangle = 0, \nonumber \\
&\langle b_a^{in}(t)b_a^{in\dagger}(t') \rangle = \delta(t-t'), \hspace*{0.5cm} \langle b_a^{in\dagger}(t)b_a^{in}(t') \rangle = 0 , \nonumber \\
&\langle b_m^{in}(t)b_m^{in\dagger}(t') \rangle = (n_{th}^j+1)\delta(t-t'), \\ 
&\langle b_m^{in\dagger}(t)b_m^{in}(t') \rangle = n_{th}^j\delta(t-t'), \nonumber \\ 
\end{eqnarray}
where $n_{th}$ is the thermal phonon occupation of the mechanical resonator defined as $n_{th}=[\rm exp(\frac{\hbar \omega_m}{k_bT})-1]^{-1}$, where $\rm k_b$ is the Boltzmann constant. Owing to the high-frequency Brillouin mode $b_a$ ($\omega_m \ll \omega_a$) the thermal acoustic occupancy has been neglected. To foster the entanglement feature in our system, we define the following quadrature operators, $\delta X_{\mathcal{O}} =\frac{\delta \mathcal{O}^\dagger + \delta \mathcal{O} }{\sqrt{2}}$, $\delta Y_{\mathcal{O}} =i\frac{\delta \mathcal{O}^\dagger - \delta \mathcal{O}}{\sqrt{2}}$, together with their related noise quadratures, $\delta X_{\mathcal{O}}^{in} =\frac{\delta \mathcal{O}^{\dagger in} + \delta \mathcal{O}^{in}}{\sqrt{2}}$, $\delta Y_{\mathcal{O}}^{in} =i\frac{\delta \mathcal{O}^{\dagger in} - \delta \mathcal{O}^{in}}{\sqrt{2}}$, where $\mathcal{O}\equiv a_1,b_a,b_m$. The quadrature dynamics can be set in its compact form, 
\begin{equation}\label{eq:quadra}
\dot{z}={\rm A} z+z^{in},
\end{equation}
where ${z}=(\delta I,\delta \phi, \delta q_a,\delta p_a, \delta q_m,\delta p_m)^T$, \\ $z^{in}=(\sqrt{\kappa} I^{in},\sqrt{\kappa} \phi^{in},\sqrt{\gamma_a} q_a^{in},\sqrt{\gamma_a} p_a^{in},\sqrt{\gamma_m} q_m^{in},\sqrt{\gamma_m} p_m^{in},)^T$ and the matrix $\rm A$ reads,
\begin{equation}\label{eq:matrix}
{\rm A}=
\begin{pmatrix}
-\frac{\kappa_1}{2}&-\tilde{\Delta}&0&-G_a&0&0 \\
\tilde{\Delta}&-\frac{\kappa_1}{2}&G_a&0&2G_m&0 \\ 
0&-G_a&-\frac{\gamma_a}{2}&\Delta_a& 0 & 0 \\
G_a&0&-\Delta_a&-\frac{\gamma_a}{2}&0 & 0 \\
0&0&0&0&-\frac{\gamma_m}{2}&\omega_m \\
2G_m&0&0&0&-\omega_m& -\frac{\gamma_m}{2}
\end{pmatrix},
\end{equation}
where we have assumed the effective couplings $G_m$ and $G_a$ real for simplicity.

\section{Steady-state optomechanical entanglement}\label{sec:entag}

To investigate the steady-state optomechanical entanglement induced by the backward stimulated
Brillouin scattering, we need to evaluate the covariance matrix of our system. The elements of this covariance matrix fulfill  $V_{ij}=\frac{\langle z_i z_j  + z_j z_i \rangle}{2}$, which also satisfy the motional equation,
\begin{equation}\label{eq:lyap1}
\dot{V}={\rm A}V+V{\rm A^T}+D,
\end{equation}
where $D$ is the diagonal diffusion matrix expressed as $D=Diag[\frac{\kappa}{2},  \frac{\kappa}{2}, \frac{\gamma_a}{2}, \frac{\gamma_a}{2}, \frac{\gamma_m}{2}(2n_{th} + 1), \frac{\gamma_m}{2}(2n_{th} + 1)]$.  To be meaningful, the matrix $\rm A$ must satisfy the Routh-Huritz stability criterion, i.e., all its eigenvalues should have negative real parts. Such condition has been fulfilled with the parameters range used in our investigation. To figure out the steady-state behavior of the entanglement, we assume that the dynamic in \autoref{eq:lyap1} is no longer time dependent, which reduces to the Lyaponuv equation,

\begin{equation}\label{eq:lyap}
{\rm A}V+V{\rm A^T}=-D.
\end{equation}
The $V_{ij}$ elements can be analytically evaluated, but this leads to a tedious task, and to cumbersome expressions. Therefore, these expressions will be computed numerically instead. In general, the covariance matrix takes the form 
\begin{equation}\label{cov}
\rm{V}=
\begin{pmatrix} 
V_{\alpha}&V_{\alpha,\beta}&V_{\alpha,\gamma} \\
V_{\alpha,\beta}^{\intercal}&V_{\beta}&V_{\beta,\gamma} \\ 
V_{\alpha,\gamma}^{\intercal}&V_{\beta,\gamma}^{\intercal}&V_{\gamma}
\end{pmatrix},
\end{equation}
where $V_i$ and  $V_{ij}$ are blocs of $2\times2$ matrices (with $i,j \equiv \alpha,\beta,\gamma$). 
The diagonal blocks denoted as $V_i$ correspond to the optical mode ($i = \alpha$), the mechanical mode ($i = \beta$), and the acoustic mode ($i = \gamma$), respectively. The off-diagonal blocks capture the correlations between different subsystems. For instance, $V_{\alpha,\beta}$ describes the correlations between the driving field and the mechanical resonators, and $V_{\alpha,\gamma}$ describes the correlations between the driving field and the acoustic mode, while $V_{\beta,\gamma}$ represents the correlations between the mechanical mode and the acoustic mode. A bipartite entanglement within two subsystems can be quantified using the logarithmic negativity ($E_N$), which can be evaluated by tracing out the non-necessary third mode. The logarithmic negativity $E_N$ is defined as,
\begin{equation}\label{eq:en}
E_N=\rm max[0,-\ln(2\nu^-)], 
\end{equation}
where $\nu^- = \frac{1}{\sqrt{2}}\sqrt{\sum(\chi)-\sqrt{\sum(\chi)^2-4\rm det\chi}}$. From \autoref{eq:en}, there is an entanglement emerging from the system if and only if the condition $\nu^-<1/2$ is fulfilled, which is equivalent to Simon's necessary and sufficient entanglement criterion for Gaussian states. The covariance matrix $\chi$, for the targeted subsystems, is defined as,
\begin{equation}
\rm{\chi}=\begin{pmatrix} 
V_i&V_{ij} \\
V_{ij}^{\intercal}&V_j
\end{pmatrix},
\end{equation}
such that $\sum(\chi)=\rm{det V_i+det V_j-2det V_{ij}}$. Throughout the work, we consider the red-sideband detuning driving for the mechanical resonator ($\tilde{\Delta}=-\omega_m$), where sufficient heating processes are suppressed in the system
\begin{figure}[tbh]
\begin{center}
  \resizebox{0.5\textwidth}{!}{
  \includegraphics{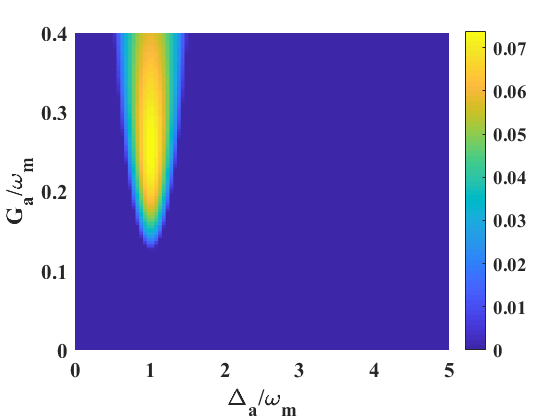}}
  \end{center}
\caption{Optomechanical entanglement in colorbar versus  $G_a$ and $\Delta_a$. The parameters used are $\omega_m/2\pi=1\rm {MHz}$, $g_m=10^{-4}\omega_m$, $\kappa=0.02\omega_m$, $\gamma_a=0.4\omega_m$, $\gamma_m=10^{-4}\omega_m$, $n_{th}=100$, $G_m=0.15\omega_m$, and $\tilde{\Delta}=-\omega_m$.}
\label{fig:Fig2}
\end{figure}

To get insight of the optomechanical entanglement in our system, we first search for optimal acoustic parameters which foster stable entanglement. These parameters include the dissipation rate ($\gamma_a$), the effective coupling ($G_a$), and the effective detuning ($\Delta_a$). The experimentally feasible parameters we have used for our investigation are $\omega_m/2\pi=1\rm {MHz}$, $g_m=10^{-4}\omega_m$, $\kappa=0.02\omega_m$, $\gamma_a=0.4\omega_m$, $\gamma_m=10^{-4}\omega_m$, $n_{th}=100$, $G_m=0.15\omega_m$, and $\tilde{\Delta}=-\omega_m$. \autoref{fig:Fig2} depicts an optomechanical entanglement in the ($G_a$,$\Delta_a$) space, where we can observe that the optimal acoustic detuning is near $\Delta_a=\omega_m$. It can be seen that the entanglement is centered around this optimal value of the detuning. With the moderated value of the optomechanical coupling $G_m\approx0.15\omega_m$, it can be observed that there is a threshold value of $G_a$ from where the mechanical and optical modes are entangled. Indeed, for $G_a\lesssim1.1\omega_m$ there is no entanglement while above this threshold, our system exhibits optomechanical entanglement. This clearly shows that the acoustic effective coupling $G_a$, which is induced through the backward stimulated Brillouin scattering process, is a key ingredient required to foster optomechanical entanglement in our proposal.   

\begin{figure}[tbh]
\begin{center}
  \resizebox{0.5\textwidth}{!}{
  \includegraphics{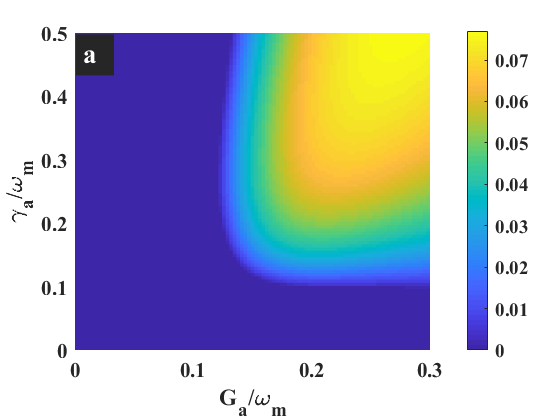}}
  \resizebox{0.5\textwidth}{!}{
  \includegraphics{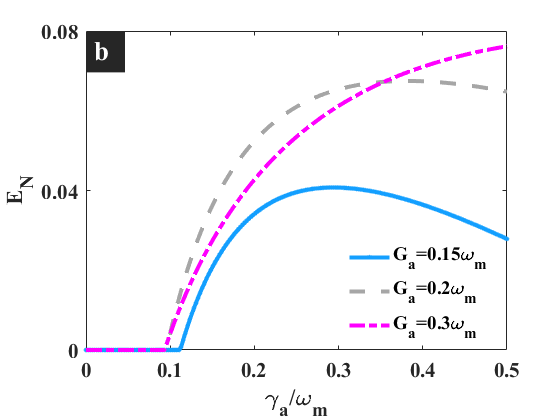}}
  \end{center}
\caption{(a) Optomechanical entanglement in colorbar versus $\gamma_a$ and $G_a$. (b) Optomechanical entanglement versus $\gamma_a$ for different values of $G_a$, i.e., $G_a=0.15\omega_m$ for the full line, $G_a=0.2\omega_m$ for the dashed line, and $G_a=0.3\omega_m$ for the dash-dotted line. The acoustic detuning has been fixed to $\Delta_a=\omega_m$, and the other parameters are the same as those in \autoref{fig:Fig2}.}
\label{fig:Fig3}
\end{figure}
With this optimal acoustic detuning in mind, we plotted the entanglement versus the acoustic parameters $\gamma_a$ and $G_a$ as shown in \autoref{fig:Fig3}a. The merit of this figure is to further reveal the key role played by the acoustic parameters in the entanglement generation process. As it can be seen, the acoustic decay rate $\gamma_a$ induces entanglement when it reaches a sort of threshold around $\gamma_a\approx0.1\omega_m$ as seen in \autoref{fig:Fig3}a. This reveals that optomechanical entanglement is generated when the acoustic decay sufficiently exceeds the optical decay ($\gamma_a\gg\kappa$). As predicted in \autoref{fig:Fig2}, \autoref{fig:Fig3}a shows that the entanglement threshold of the acoustic coupling is still near $G_a=0.15\omega_m$. From these thresholds of the acoustic parameters, \autoref{fig:Fig3}a shows how the entanglement becomes robust as the strength of both $G_a$ and $\gamma_a$ increases. As shown in \autoref{fig:Fig3}b, such behavior is reinforced when the strength of $G_a$ and $\gamma_a$ are strong enough. Indeed, \autoref{fig:Fig3}b displays the entanglement versus the decay rate $\gamma_a$ for different values of $G_a$. In can be seen that the dashed line (for $G_a=0.2\omega_m$) provides stronger entanglement compared to the full line, where $G_a=0.15\omega_m$. For the dashed line ($G_a=0.2\omega_m$) and the dash-dotted line (for $G_a=0.3\omega_m$), however, stronger entanglement is generated by the dash-dotted line only for a large enough decay rate $\gamma_a$. The feature depicted in \autoref{fig:Fig3} highlights the fact that more optomechanical entanglement requires large strength of the acoustic decay rate in our proposal.   

\begin{figure}[tbh]
  \begin{center}
  \resizebox{0.5\textwidth}{!}{
  \includegraphics{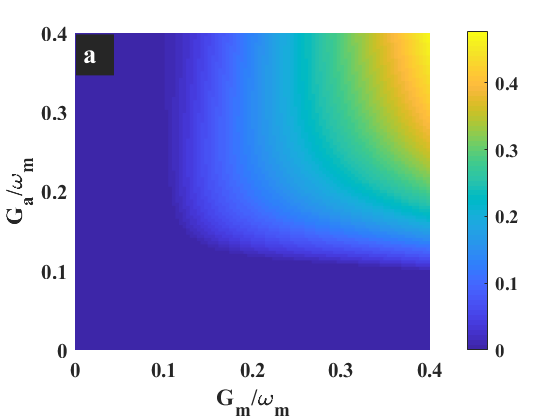}}
  \resizebox{0.5\textwidth}{!}{
  \includegraphics{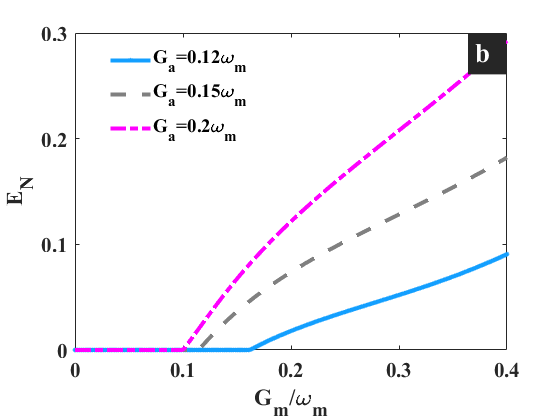}}
  \end{center}
\caption{(a) Optomechanical entanglement (in colorbar) versus $G_m$ and $G_m$. (b) The same entanglement vesus $G_m$ for specific values of $G_a$. The full, dashed, and dash-dotted lines correspond to $G_a=0.12\omega_m$, $G_a=0.15\omega_m$, and $G_a=0.2\omega_m$, respectively. We have set $\Delta_a=\omega_m$, and the other parameters are the same as those in \autoref{fig:Fig2}.}
\label{fig:Fig4}
\end{figure}

From the above discussion regarding the optomechanical generation assisted by acoustic parameters induced via BSBS process, we aim now to enhance this entanglement through the optomechanical coupling as it is usually carried out. Indeed, the entanglement shown in \autoref{fig:Fig2} and \autoref{fig:Fig3} corresponds to a moderate value of the optomechanical coupling ($G_m=0.15\omega_m$), and for a thermal occupancy of $n_{th}=100$. It is worth to mention that even in such a moderate choice of parameter case, a sufficient optomechanical entanglement is generated (see colorbars). In order to enhance this entanglement, we displayed in \autoref{fig:Fig4}a the entanglement versus mechanical ($G_m$) and acoustic ($G_a$) couplings. As aforementioned in the previous figures, \autoref{fig:Fig4}a shows how the threshold of the acoustic coupling still appears near $G_a=0.15\omega_m$, meaning that no matter is the strength of the mechanical coupling, optomechanical entanglement requires the BSBS process in our proposal. This feature exemplifies the main role played by the BSBS in our investigation. Once that this threshold is reached, \autoref{fig:Fig4}a shows how robust entanglement is induced as the mechanical coupling ($G_m$) increases. This amount of the enhanced entanglement can be appreciated by comparing the colorbar of \autoref{fig:Fig4}a to those of \autoref{fig:Fig2} and \autoref{fig:Fig3}. To further point out this enhancement feature, we display in \autoref{fig:Fig4}b the logarithmic negativity versus the mechanical coupling for different values of $G_a$. Once the threshold is reached, \autoref{fig:Fig4}b shows how the robust entanglement is generated as the strength of $G_a$ increases. Indeed, the full line ($G_a=0.12\omega_m$) generates less entanglement than the dashed line ($G_a=0.15\omega_m$), and the entanglement induced for $G_a=0.2\omega_m$ (dash-dotted line) is stronger compared with the one from the dashed line ($G_a=0.15\omega_m$). From the above discussion, it can be stated that the BSBS process induces optomechanical entanglement in our proposal, and the resulted entanglement can be enhanced through the mechanical coupling.
    
\begin{figure}[tbh]
  \begin{center}
  \resizebox{0.5\textwidth}{!}{
  \includegraphics{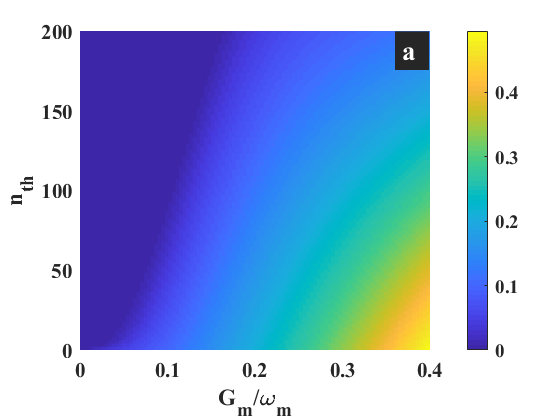}}
  \resizebox{0.5\textwidth}{!}{
  \includegraphics{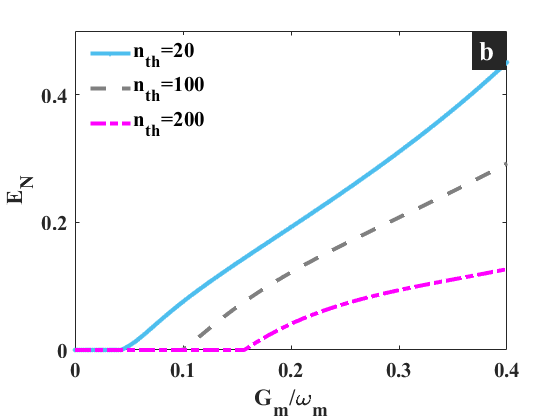}}
  \end{center}
\caption{(a) Optomechanical entanglement (in colorbar) versus the thermal occupancy $n_{th}$ and $G_m$. (b) The same entanglement vesus $G_m$ for specific values of $n_{th}$. The full, dashed, and dash-dotted lines correspond to $n_{th}=20$, $n_{th}=100$, and $n_{th}=200$, respectively. We have set $\Delta_a=\omega_m$, $G_a=0.2\omega_m$, and the other parameters are the same as those in \autoref{fig:Fig2}.}
\label{fig:Fig5}
\end{figure}

As our system couples to the mechanical bath, the resulting interaction adds additional heating channels to the dynamic, which are not fully suppressed by the red-sideband pumping process. Such interaction that hinders the entanglement generation requires special attention to be paid in order to point out the thermal noise effect. For this purpose, \autoref{fig:Fig5}a displays the entanglement behavior versus the thermal occupancy $n_{th}$ and the mechanical coupling $G_m$. It can be seen that there is no entanglement for weak mechanical coupling ($G_m\lesssim 0.05\omega_m$) even when there is no mechanical bath interacting with the system ($n_{th}=0$). However, this entanglement grows fast and reaches large values for weak thermal phonon occupancy ($n_{th}\sim20$). As the phonon occupancy $n_{th}$ increases, one observes a weakness of the entanglement. From $G_m>0.15\omega_m$, it can be seen that the entanglement resists against thermal occupancy for up to $n_{th}=200$. To highlights this robustness of the entanglement against $n_{th}$, \autoref{fig:Fig5}b depicts the behavior of the generated entanglement versus $G_m$ for different values of $n_{th}$. It can be seen that the generated entanglement is large for weak interaction with the bath (full line), and the logarithmic negativity decreases as the phonon occupancy increases (see dashed and dash-dotted lines). Moreover, this entanglement survives for phonon number up to $n_{th}=200$ (dash-dotted lines), revealing the robustness of the generated entanglement against thermal fluctuation.     

\section{Conclusion} \label{sec:concl}
This work has investigated on a generation of steady state optomechanical entanglement through a backward stimulated Brillouin scattering (BSBS) process. The benchmark system is made of an acoustic mode that couples to two optical modes through the BSBS process, and a mechanical oscillator that couples to the same optical modes via the standard optomechanical radiation pressure. For a moderated value of the effective mechanical coupling, the optomechanical generated requires acoustic parameters with specific threshold. Moreover, the induced entanglement is relatively weak in general. As the mechanical coupling strength increases, the value of the induced entanglement is enhanced at least up to one order of magnitude. The effect of the thermal noise has been carried out, which reveals how the generated entanglement is robust enough against thermal fluctuation. Our work provides a new scheme based on BSBS to generate robust entanglement in optomechanical systems, and this can be extended to microwaves and hybrid optomechanical structures.

\section*{Acknowledgments}
\textbf{Funding:} This work has been carried out under the Iso-Lomso Fellowship at Stellenbosch Institute for Advanced Study (STIAS), Wallenberg Research Centre at Stellenbosch University, Stellenbosch 7600, South Africa. K.S. Nisar is grateful to the funding from Prince Sattam bin Abdulaziz University, Saudi Arabia project number (PSAU/2024/R/1445). The authors are thankful to the Deanship of Graduate Studies and Scientific Research at University of Bisha for supporting this work through the Fast-Track Research Support Program.

\textbf{Competing Interests:} All authors declare no competing interests.

\section*{Data Availability}
Relevant data are included in the manuscript
and supporting information. Supplement data are available upon reasonable request.

 \appendix

\section{Effective Hamiltonian}\label{App}
The Hamiltonian of our proposal in the natural frame is given by,
\begin{align}
H&=\sum_{j=1,2} \omega_{c_j} a_j^\dagger  a_j + \omega_a b_a^\dagger b_a + \omega_m b_m^\dagger b_m  \\&- \sum_{j=1,2}g_{m}a_j^\dagger a_j   (b_m +b_m^\dagger)\\&+\sum_{j=1,2}iE_j(a_j^{\dagger }e^{-i\omega_{p_j} t} - a_je^{i\omega_{p_j} t})- g_a( a_1^\dagger a_2 b_a+ a_1 a_2^\dagger b_a^\dagger).
\end{align}
By moving to the frame rotating with the frequency $\omega_{p_1}a_1^\dagger a_1 +\omega_{p_2} a_2^\dagger a_2 +({\omega_{p_1}-\omega_{p_2}})b_a^\dagger b_a $, the above Hamiltonian becomes,

\begin{align}\label{eq:hamil}
H'&=-\sum_{j=1,2} \Delta_{j} a_j^\dagger  a_j +  \Delta_a b_a^\dagger b_a+\omega_m b_m^\dagger b_m \\&- \sum_{j=1,2}g_{m}a_j^\dagger a_j (b_m +b_m^\dagger)\\&+iE_1(a_1^{\dagger } - a_1) +iE_2(a_2^{\dagger} - a_2) - g_a( a_1^\dagger a_2 b_a+ a_1 a_2^\dagger b_a^\dagger),
\end{align}
with $\Delta_j=\omega_{p_j}-\omega_{c_j}$, and $\Delta_a=\omega_a+\omega_{p_2}-\omega_{p_1}$. 
By considering that the control field $a_2$ is strong enough compared to $a_1$, it can be treated classically by deriving its steady-state as,
\begin{align}
\dot{a}_2 &= i[H',a_2], \\
 &=(i\Delta_2-\frac{\kappa_2}{2})a_2 + ig_{m}a_2(b_m+b_m^\dagger) + E_2 + ig_a a_1 b_a^\dagger \nonumber \\
 &=(i\Delta'_2-\frac{\kappa_2}{2})a_2 + E_2 + ig_a a_1 b_a^\dagger,
\end{align}
with $\Delta'_2=\Delta_2 + g_{m}a_2(b_m+b_m^\dagger)$. The steady-state solution ($\dot{a}_2=0$) yields, 
\begin{align}
\alpha_{2}&\sim\frac{-E_2}{i\Delta'_2-\frac{\kappa_2}{2}} \hspace{1em} \text{or} \hspace{1em} |\alpha_{2}|\sim\frac{E_2}{\sqrt{\Delta_2^{'^2}+\frac{\kappa_2^2}{4}}}.
\end{align}
By replacing this expression in the rest of the Hamiltonian in \autoref{eq:hamil}, we get the following reduced Hamiltonian,
\begin{align}
H'&=-\Delta_{1} a_1^\dagger  a_1 +  \Delta_a b_a^\dagger b_a+\omega_m b_m^\dagger b_m- g_{m} a_1^\dagger a_1 (b_m +b_m^\dagger)\nonumber \\ &+iE_1(a_1^{\dagger } - a_1)- {G_a} ( a_1^\dagger b_a + a_1 b_a^\dagger ),
\end{align}
where the acoustic effective coupling is $G_a=g_a\alpha_{2}$ as mentioned in the main text.

\printcredits

\bibliographystyle{unsrtnat}

\bibliography{Entanglement}

\begin{thebibliography}{42}
\providecommand{\natexlab}[1]{#1}
\providecommand{\url}[1]{\texttt{#1}}
\expandafter\ifx\csname urlstyle\endcsname\relax
  \providecommand{\doi}[1]{doi: #1}\else
  \providecommand{\doi}{doi: \begingroup \urlstyle{rm}\Url}\fi

\bibitem[Walter and Marquardt(2016)]{Walter_2016}
Stefan Walter and Florian Marquardt.
\newblock Classical dynamical gauge fields in optomechanics.
\newblock \emph{New Journal of Physics}, 18\penalty0 (11):\penalty0 113029,
  November 2016.
\newblock ISSN 1367-2630.
\newblock \doi{10.1088/1367-2630/18/11/113029}.

\bibitem[Pokharel et~al.(2022)Pokharel, Xu, Venkatachalam, Collin, and
  Zhou]{Pokharel_2022}
Alok Pokharel, Hao Xu, Srisaran Venkatachalam, Eddy Collin, and Xin Zhou.
\newblock Coupling capacitively distinct mechanical resonators for
  room-temperature phonon-cavity electromechanics.
\newblock \emph{Nano Letters}, 22\penalty0 (18):\penalty0 7351--7357, September
  2022.
\newblock ISSN 1530-6992.
\newblock \doi{10.1021/acs.nanolett.2c01848}.

\bibitem[Foulla et~al.(2017)Foulla, Djorw{\'{e}}, Kingni, and
  Engo]{Foulla_2017}
D.~Platou Foulla, P.~Djorw{\'{e}}, S.~Takougang Kingni, and S.~G.~Nana Engo.
\newblock Optomechanical systems close to the conservative limit.
\newblock \emph{Physical Review A}, 95\penalty0 (1):\penalty0 123, jan 2017.
\newblock \doi{10.1103/PhysRevA.95.013831}.

\bibitem[Djorwe et~al.(2022)Djorwe, Yves Effa, and G. Nana Engo]{Djor.2022}
Philippe Djorwe, Joseph Yves Effa, and S.~G. Nana Engo.
\newblock Hidden attractors and metamorphoses of basin boundaries in
  optomechanics.
\newblock \emph{Nonlinear Dynamics}, 111\penalty0 (6):\penalty0 5905--5917,
  December 2022.
\newblock ISSN 1573-269X.
\newblock \doi{10.1007/s11071-022-08139-2}.

\bibitem[Aspelmeyer et~al.(2010)Aspelmeyer, Gröblacher, Hammerer, and
  Kiesel]{Aspelmeyer_2010}
M.~Aspelmeyer, S.~Gröblacher, K.~Hammerer, and N.~Kiesel.
\newblock Quantum optomechanics—throwing a glance [invited].
\newblock \emph{Journal of the Optical Society of America B}, 27\penalty0
  (6):\penalty0 A189, May 2010.
\newblock ISSN 1520-8540.
\newblock \doi{10.1364/JOSAB.27.00A189}.

\bibitem[Lemonde et~al.(2016)Lemonde, Didier, and Clerk]{Lemonde.2016}
Marc-Antoine Lemonde, Nicolas Didier, and Aashish~A. Clerk.
\newblock Enhanced nonlinear interactions in quantum optomechanics via
  mechanical amplification.
\newblock \emph{Nature Communications}, 7\penalty0 (1):\penalty0 11338, April
  2016.
\newblock ISSN 2041-1723.
\newblock URL \url{https://doi.org/10.1038/ncomms11338}.

\bibitem[Djorw{\'{e}} et~al.(2012)Djorw{\'{e}}, Mb{\'{e}}, Engo, and
  Woafo]{Djor.2012}
P.~Djorw{\'{e}}, J.~H.~Talla Mb{\'{e}}, S.~G.~Nana Engo, and P.~Woafo.
\newblock Nonlinearity-induced limitations on cooling in optomechanical
  systems.
\newblock \emph{Physical Review A}, 86\penalty0 (4):\penalty0 043816, oct 2012.
\newblock \doi{10.1103/PhysRevA.86.043816}.

\bibitem[Li et~al.(2022)Li, Zhou, Wan, Zhang, Shen, Li, Zou, Guo, and
  Dong]{Li_2022}
Jin Li, Zhong-Hao Zhou, Shuai Wan, Yan-Lei Zhang, Zhen Shen, Ming Li,
  Chang-Ling Zou, Guang-Can Guo, and Chun-Hua Dong.
\newblock All-optical synchronization of remote optomechanical systems.
\newblock \emph{Physical Review Letters}, 129\penalty0 (6):\penalty0 063605,
  August 2022.
\newblock ISSN 1079-7114.
\newblock \doi{10.1103/PhysRevLett.129.063605}.

\bibitem[Colombano et~al.(2019)Colombano, Arregui, Capuj, Pitanti, Maire,
  Griol, Garrido, Martinez, Sotomayor-Torres, and
  Navarro-Urrios]{Colombano_2019}
M.~F. Colombano, G.~Arregui, N.~E. Capuj, A.~Pitanti, J.~Maire, A.~Griol,
  B.~Garrido, A.~Martinez, C.~M. Sotomayor-Torres, and D.~Navarro-Urrios.
\newblock Synchronization of optomechanical nanobeams by mechanical
  interaction.
\newblock \emph{Physical Review Letters}, 123\penalty0 (1):\penalty0 017402,
  July 2019.
\newblock ISSN 1079-7114.
\newblock \doi{10.1103/PhysRevLett.123.017402}.

\bibitem[Djorw{\'{e}} et~al.(2020)Djorw{\'{e}}, Pennec, and
  Djafari-Rouhani]{Djorwe.2020}
P.~Djorw{\'{e}}, Y.~Pennec, and B.~Djafari-Rouhani.
\newblock Self-organized synchronization of mechanically coupled resonators
  based on optomechanics gain-loss balance.
\newblock \emph{Physical Review B}, 102\penalty0 (15):\penalty0 155410, oct
  2020.
\newblock \doi{10.1103/PhysRevB.102.155410}.

\bibitem[Roque et~al.(2020)Roque, Marquardt, and Yevtushenko]{Roque.2020}
Thales~Figueiredo Roque, Florian Marquardt, and Oleg~M Yevtushenko.
\newblock Nonlinear dynamics of weakly dissipative optomechanical systems.
\newblock \emph{New Journal of Physics}, 22\penalty0 (1):\penalty0 013049,
  January 2020.
\newblock ISSN 1367-2630.
\newblock \doi{10.1088/1367-2630/ab6522}.

\bibitem[Xu et~al.(2024)Xu, Zhu, Chen, Li, and Zhang]{Xu_2024}
Xiangming Xu, Huatao Zhu, Shuwen Chen, Feiyu Li, and Xin Zhang.
\newblock Nonlinear dynamics of cavity optomechanical-thermal systems.
\newblock \emph{Optics Express}, 32\penalty0 (5):\penalty0 7611, February 2024.
\newblock ISSN 1094-4087.
\newblock \doi{10.1364/OE.515095}.

\bibitem[Navarro-Urrios et~al.(2017)Navarro-Urrios, Capuj, Colombano, García,
  Sledzinska, Alzina, Griol, Martínez, and Sotomayor-Torres]{Navarro.2017}
Daniel Navarro-Urrios, Néstor~E. Capuj, Martín~F. Colombano, P.~David
  García, Marianna Sledzinska, Francesc Alzina, Amadeu Griol, Alejandro
  Martínez, and Clivia~M. Sotomayor-Torres.
\newblock Nonlinear dynamics and chaos in an optomechanical beam.
\newblock \emph{Nature Communications}, 8\penalty0 (1):\penalty0 14965, April
  2017.
\newblock ISSN 2041-1723.
\newblock \doi{10.1038/ncomms14965}.

\bibitem[Zhu et~al.(2023)Zhu, Hu, Wu, and Lü]{Zhu_2023}
Gui-Lei Zhu, Chang-Sheng Hu, Ying Wu, and Xin-You Lü.
\newblock Cavity optomechanical chaos.
\newblock \emph{Fundamental Research}, 3\penalty0 (1):\penalty0 63--74, January
  2023.
\newblock ISSN 2667-3258.
\newblock \doi{10.1016/j.fmre.2022.07.012}.

\bibitem[Stella et~al.(2023)Stella, Djorwe, Murielle, and
  Serge~Guy]{Stella_2023}
Mbokop~Tchounda Stella, Philippe Djorwe, Tchakui Murielle, and Nana~Engo
  Serge~Guy.
\newblock Chaos control and exceptional point engineering via dissipative
  optomechanical coupling.
\newblock \emph{Physica Scripta}, 99:\penalty0 025215, December 2023.
\newblock ISSN 1402-4896.
\newblock \doi{10.1088/1402-4896/ad195c}.

\bibitem[Rodrigues et~al.(2021)Rodrigues, Kersul, Primo, Lipson, Alegre, and
  Wiederhecker]{Rodrigues_2021}
Caique~C. Rodrigues, Cauê~M. Kersul, André~G. Primo, Michal Lipson, Thiago
  P.~Mayer Alegre, and Gustavo~S. Wiederhecker.
\newblock Optomechanical synchronization across multi-octave frequency spans.
\newblock \emph{Nature Communications}, 12\penalty0 (1):\penalty0 5625,
  September 2021.
\newblock ISSN 2041-1723.
\newblock \doi{10.1038/s41467-021-25884-x}.

\bibitem[Berra et~al.(2023)Berra, Agnesi, Stanco, Avesani, Kuklewski, Matter,
  Vallone, and Villoresi]{Berra_2023}
Federico Berra, Costantino Agnesi, Andrea Stanco, Marco Avesani, Michal
  Kuklewski, Daniel Matter, Giuseppe Vallone, and Paolo Villoresi.
\newblock Synchronization of quantum communications over an optical classical
  communications channel.
\newblock \emph{Applied Optics}, 62\penalty0 (30):\penalty0 7994, October 2023.
\newblock ISSN 2155-3165.
\newblock \doi{10.1364/AO.500416}.

\bibitem[Madiot et~al.(2022)Madiot, Correia, Barbay, and Braive]{Madiot_2022}
Guilhem Madiot, Franck Correia, Sylvain Barbay, and Remy Braive.
\newblock Random number generation with a chaotic electromechanical resonator.
\newblock \emph{Nanotechnology}, 33\penalty0 (47):\penalty0 475204, September
  2022.
\newblock ISSN 1361-6528.
\newblock \doi{10.1088/1361-6528/ac86da}.

\bibitem[Wu et~al.(2019)Wu, Flor~Flores, Bai, Yang, Xiong, Yu, Lo, Kwong, Duan,
  and Wong]{Wu_2019}
Jiagui Wu, Jaime~G. Flor~Flores, Qingsong Bai, Jinghui Yang, Xueyan Xiong,
  Mingbin Yu, Guoqiang Lo, Dim-Lee Kwong, Shukai Duan, and Chee~Wei Wong.
\newblock Dynamical chaos in silicon cavity optomechanics for
  physically-encrypted secure communications.
\newblock \penalty0 (1-2), 2019.
\newblock \doi{10.1364/CLEO_SI.2019.SF1J.4}.

\bibitem[Barzanjeh et~al.(2021)Barzanjeh, Xuereb, Gröblacher, Paternostro,
  Regal, and Weig]{Barzanjeh_2021}
Shabir Barzanjeh, André Xuereb, Simon Gröblacher, Mauro Paternostro, Cindy~A.
  Regal, and Eva~M. Weig.
\newblock Optomechanics for quantum technologies.
\newblock \emph{Nature Physics}, 18\penalty0 (1):\penalty0 15--24, December
  2021.
\newblock ISSN 1745-2481.
\newblock \doi{10.1038/s41567-021-01402-0}.

\bibitem[Qin et~al.(2022)Qin, Miranowicz, and Nori]{Qin_2022}
Wei Qin, Adam Miranowicz, and Franco Nori.
\newblock Beating the 3 db limit for intracavity squeezing and its application
  to nondemolition qubit readout.
\newblock \emph{Physical Review Letters}, 129\penalty0 (12):\penalty0 123602,
  September 2022.
\newblock ISSN 1079-7114.
\newblock \doi{10.1103/PhysRevLett.129.123602}.

\bibitem[Banerjee et~al.(2023)Banerjee, Kalita, and Sarma]{Banerjee_2023}
Priyankar Banerjee, Sampreet Kalita, and Amarendra~K. Sarma.
\newblock Robust mechanical squeezing beyond 3db in a quadratically coupled
  optomechanical system.
\newblock \emph{Journal of the Optical Society of America B}, 40\penalty0
  (6):\penalty0 1398, May 2023.
\newblock ISSN 1520-8540.
\newblock \doi{10.1364/JOSAB.483944}.

\bibitem[Riedinger et~al.(2018)Riedinger, Wallucks, Marinković, Löschnauer,
  Aspelmeyer, Hong, and Gröblacher]{Riedinger_2018}
Ralf Riedinger, Andreas Wallucks, Igor Marinković, Clemens Löschnauer, Markus
  Aspelmeyer, Sungkun Hong, and Simon Gröblacher.
\newblock Remote quantum entanglement between two micromechanical oscillators.
\newblock \emph{Nature}, 556\penalty0 (7702):\penalty0 473--477, April 2018.
\newblock ISSN 1476-4687.
\newblock \doi{10.1038/s41586-018-0036-z}.

\bibitem[Kotler et~al.(2021)Kotler, Peterson, Shojaee, Lecocq, Cicak,
  Kwiatkowski, Geller, Glancy, Knill, Simmonds, Aumentado, and
  Teufel]{Kotler_2021}
Shlomi Kotler, Gabriel~A. Peterson, Ezad Shojaee, Florent Lecocq, Katarina
  Cicak, Alex Kwiatkowski, Shawn Geller, Scott Glancy, Emanuel Knill,
  Raymond~W. Simmonds, José Aumentado, and John~D. Teufel.
\newblock Direct observation of deterministic macroscopic entanglement.
\newblock \emph{Science}, 372\penalty0 (6542):\penalty0 622--625, May 2021.
\newblock ISSN 1095-9203.
\newblock \doi{10.1126/science.abf2998}.

\bibitem[Riedinger et~al.(2016)Riedinger, Hong, Norte, Slater, Shang, Krause,
  Anant, Aspelmeyer, and Gröblacher]{Riedin.2016}
Ralf Riedinger, Sungkun Hong, Richard~A. Norte, Joshua~A. Slater, Juying Shang,
  Alexander~G. Krause, Vikas Anant, Markus Aspelmeyer, and Simon Gröblacher.
\newblock Non-classical correlations between single photons and phonons from a
  mechanical oscillator.
\newblock \emph{Nature}, 530\penalty0 (7590):\penalty0 313--316, January 2016.
\newblock ISSN 1476-4687.
\newblock \doi{10.1038/nature16536}.

\bibitem[Purdy et~al.(2017)Purdy, Grutter, Srinivasan, and Taylor]{Purdy_2017}
T.~P. Purdy, K.~E. Grutter, K.~Srinivasan, and J.~M. Taylor.
\newblock Quantum correlations from a room-temperature optomechanical cavity.
\newblock \emph{Science}, 356\penalty0 (6344):\penalty0 1265--1268, June 2017.
\newblock ISSN 1095-9203.
\newblock \doi{10.1126/science.aag1407}.

\bibitem[Mari et~al.(2013)Mari, Farace, Didier, Giovannetti, and
  Fazio]{Mari_2013}
A.~Mari, A.~Farace, N.~Didier, V.~Giovannetti, and R.~Fazio.
\newblock Measures of quantum synchronization in continuous variable systems.
\newblock \emph{Physical Review Letters}, 111\penalty0 (10):\penalty0 103605,
  September 2013.
\newblock ISSN 1079-7114.
\newblock \doi{10.1103/PhysRevLett.111.103605}.

\bibitem[Garg et~al.(2023)Garg, Manju, Dasgupta, and Biswas]{Garg_2023}
Devender Garg, Manju, Shubhrangshu Dasgupta, and Asoka Biswas.
\newblock Quantum synchronization and entanglement of indirectly coupled
  mechanical oscillators in cavity optomechanics: A numerical study.
\newblock \emph{Physics Letters A}, 457:\penalty0 128557, January 2023.
\newblock ISSN 0375-9601.
\newblock \doi{10.1016/j.physleta.2022.128557}.

\bibitem[Zidan et~al.(2021)Zidan, Eleuch, and Abdel-Aty]{Zidan_2021}
Mohammed Zidan, Hichem Eleuch, and Mahmoud Abdel-Aty.
\newblock Non-classical computing problems: Toward novel type of quantum
  computing problems.
\newblock \emph{Results in Physics}, 21:\penalty0 103536, February 2021.
\newblock ISSN 2211-3797.
\newblock \doi{10.1016/j.rinp.2020.103536}.

\bibitem[Meher and Sivakumar(2022)]{Meher_2022}
Nilakantha Meher and S.~Sivakumar.
\newblock A review on quantum information processing in cavities.
\newblock \emph{The European Physical Journal Plus}, 137\penalty0 (8):\penalty0
  985, August 2022.
\newblock ISSN 2190-5444.
\newblock \doi{10.1140/epjp/s13360-022-03172-x}.

\bibitem[Xia et~al.(2023)Xia, Agrawal, Pluchar, Brady, Liu, Zhuang, Wilson, and
  Zhang]{Xia_2023}
Yi~Xia, Aman~R. Agrawal, Christian~M. Pluchar, Anthony~J. Brady, Zhen Liu,
  Quntao Zhuang, Dalziel~J. Wilson, and Zheshen Zhang.
\newblock Entanglement-enhanced optomechanical sensing.
\newblock \emph{Nature Photonics}, 17\penalty0 (6):\penalty0 470--477, April
  2023.
\newblock ISSN 1749-4893.
\newblock \doi{10.1038/s41566-023-01178-0}.

\bibitem[Brady et~al.(2023)Brady, Chen, Xia, Manley, Dey~Chowdhury, Xiao, Liu,
  Harnik, Wilson, Zhang, and Zhuang]{Brady_2023}
Anthony~J. Brady, Xin Chen, Yi~Xia, Jack Manley, Mitul Dey~Chowdhury, Kewen
  Xiao, Zhen Liu, Roni Harnik, Dalziel~J. Wilson, Zheshen Zhang, and Quntao
  Zhuang.
\newblock Entanglement-enhanced optomechanical sensor array with application to
  dark matter searches.
\newblock \emph{Communications Physics}, 6\penalty0 (1):\penalty0 237,
  September 2023.
\newblock ISSN 2399-3650.
\newblock \doi{10.1038/s42005-023-01357-z}.

\bibitem[Fiaschi et~al.(2021)Fiaschi, Hensen, Wallucks, Benevides, Li, Alegre,
  and Gröblacher]{Fiaschi_2021}
Niccolò Fiaschi, Bas Hensen, Andreas Wallucks, Rodrigo Benevides, Jie Li,
  Thiago P.~Mayer Alegre, and Simon Gröblacher.
\newblock Optomechanical quantum teleportation.
\newblock \emph{Nature Photonics}, 15\penalty0 (11):\penalty0 817--821, October
  2021.
\newblock ISSN 1749-4893.
\newblock \doi{10.1038/s41566-021-00866-z}.

\bibitem[Zhang et~al.(2017)Zhang, Zeng, and Chen]{Zhang_2017}
Jian-Song Zhang, Wei Zeng, and Ai-Xi Chen.
\newblock Effects of cross-kerr coupling and parametric nonlinearity on normal
  mode splitting, cooling, and entanglement in optomechanical systems.
\newblock \emph{Quantum Information Processing}, 16\penalty0 (6), May 2017.
\newblock ISSN 1573-1332.
\newblock \doi{10.1007/s11128-017-1614-y}.

\bibitem[Chakraborty and Sarma(2017)]{Cha_2017}
Subhadeep Chakraborty and Amarendra~K. Sarma.
\newblock Enhancing quantum correlations in an optomechanical system via
  cross-kerr nonlinearity.
\newblock \emph{Journal of the Optical Society of America B}, 34\penalty0
  (7):\penalty0 1503, June 2017.
\newblock ISSN 1520-8540.
\newblock \doi{10.1364/JOSAB.34.001503}.

\bibitem[Chakraborty and Sarma(2019)]{Chakrabo.2019}
Subhadeep Chakraborty and Amarendra~K. Sarma.
\newblock Delayed sudden death of entanglement at exceptional points.
\newblock \emph{Physical Review A}, 100\penalty0 (6):\penalty0 063846, December
  2019.
\newblock ISSN 2469-9934.
\newblock \doi{10.1103/PhysRevA.100.063846}.

\bibitem[Lai et~al.(2022{\natexlab{a}})Lai, Chen, Qin, Miranowicz, and
  Nori]{Lai.2022}
Deng-Gao Lai, Ye-Hong Chen, Wei Qin, Adam Miranowicz, and Franco Nori.
\newblock Tripartite optomechanical entanglement via optical-dark-mode control.
\newblock \emph{Physical Review Research}, 4\penalty0 (3), August
  2022{\natexlab{a}}.
\newblock ISSN 2643-1564.
\newblock \doi{10.1103/PhysRevResearch.4.033112}.

\bibitem[Lai et~al.(2022{\natexlab{b}})Lai, Liao, Miranowicz, and
  Nori]{Lai_2022}
Deng-Gao Lai, Jie-Qiao Liao, Adam Miranowicz, and Franco Nori.
\newblock Noise-tolerant optomechanical entanglement via synthetic magnetism.
\newblock \emph{Physical Review Letters}, 129\penalty0 (6):\penalty0 063602,
  aug 2022{\natexlab{b}}.
\newblock \doi{10.1103/PhysRevLett.129.063602}.

\bibitem[Bahl et~al.(2013)Bahl, Kim, Lee, Liu, Fan, and Carmon]{Bahl_2013}
Gaurav Bahl, Kyu~Hyun Kim, Wonsuk Lee, Jing Liu, Xudong Fan, and Tal Carmon.
\newblock Brillouin cavity optomechanics with microfluidic devices.
\newblock \emph{Nature Communications}, 4\penalty0 (1):\penalty0 1994, June
  2013.
\newblock ISSN 2041-1723.
\newblock \doi{10.1038/ncomms2994}.

\bibitem[He et~al.(2020)He, Harris, Baker, Sawadsky, Sfendla, Sachkou,
  Forstner, and Bowen]{He_2020}
Xin He, Glen~I. Harris, Christopher~G. Baker, Andreas Sawadsky, Yasmine~L.
  Sfendla, Yauhen~P. Sachkou, Stefan Forstner, and Warwick~P. Bowen.
\newblock Strong optical coupling through superfluid brillouin lasing.
\newblock \emph{Nature Physics}, 16\penalty0 (4):\penalty0 417--421, February
  2020.
\newblock ISSN 1745-2481.
\newblock \doi{10.1038/s41567-020-0785-0}.

\bibitem[Doeleman et~al.(2023)Doeleman, Schatteburg, Benevides, Vollenweider,
  Macri, and Chu]{Doeleman_2023}
H.~M. Doeleman, T.~Schatteburg, R.~Benevides, S.~Vollenweider, D.~Macri, and
  Y.~Chu.
\newblock Brillouin optomechanics in the quantum ground state.
\newblock \emph{Physical Review Research}, 5\penalty0 (4):\penalty0 043140,
  November 2023.
\newblock ISSN 2643-1564.
\newblock \doi{10.1103/PhysRevResearch.5.043140}.

\bibitem[Wang et~al.(2024)Wang, Hu, Lao, Wang, Jin, Zhou, Lei, Wang, Liu, Yang,
  and Li]{Wang_2024}
Min Wang, Zhi-Gang Hu, Chenghao Lao, Yuanlei Wang, Xing Jin, Xin Zhou, Yuechen
  Lei, Ze~Wang, Wenjing Liu, Qi-Fan Yang, and Bei-Bei Li.
\newblock Taming brillouin optomechanics using supermode microresonators.
\newblock \emph{Physical Review X}, 14\penalty0 (1):\penalty0 011056, March
  2024.
\newblock ISSN 2160-3308.
\newblock \doi{10.1103/PhysRevX.14.011056}.

\end{thebibliography}

\end{document}